\documentclass[12pt,preprint]{aastex}
\usepackage{natbib}
\usepackage{epsf}
\usepackage{graphicx,rotate}

\newcommand{\vm}{V_{\rm max}}
\newcommand{\dmb}{\Delta M_B}

\shorttitle{Evolution of distant field spirals}
\shortauthors{B\"ohm \& Ziegler}

\begin{document}

\title{Evolution of field spiral galaxies up to redshifts $z=1$
\altaffilmark{1}}
\altaffiltext{1}{Based on observations collected at the European
Southern Observatory, Cerro Paranal, Chile 
(ESO Nos. 65.O-0049, 66.A-0547, 68.A-0013, 69.B-0278B and 70.B-0251A)
and observations with the NASA/ESA \emph{Hubble Space Telescope},
PID 9502 and 9908.}


\author{Asmus B\"ohm\altaffilmark{2,3} and Bodo~L.~Ziegler\altaffilmark{2,4}}
\affil{$^2$ Institut f\"ur Astrophysik G\"ottingen, Friedrich-Hund-Platz 1, 37077
G\"ottingen, Germany}
\email{boehm@astro.physik.uni-goettingen.de, 
bziegler@astro.physik.uni-goettingen.de}
\affil{$^3$ Astrophysikalisches Institut Potsdam, An der Sternwarte 16, 14482
  Potsdam, Germany}
\email{aboehm@aip.de}
\affil{$^4$ Argelander-Institut f\"ur Astronomie, Auf dem H\"ugel 71, 53121
  Bonn, Germany}
\email{bziegler@astro.uni-bonn.de}

\begin{abstract}

We have gained intermediate--resolution spectroscopy with the FORS instruments
of the Very Large Telescope and high--resolution imaging with the Advanced Camera
for Surveys aboard \emph{HST} of a sample of 
220 distant field spiral galaxies within the FORS Deep Field
and William Herschel Deep Field. 
Spatially resolved rotation curves were extracted and fitted with synthetic
velocity fields that take into account all geometric and observational effects,
like blurring due to the slit width and seeing influence. 
Using these fits, the maximum rotation velocity $\vm$ could be determined for
124 galaxies that cover the redshift range $0.1<z<1.0$ and comprise a variety
of morphologies from early--type spirals to very late--types and irregulars. 
The luminosity--rotation velocity distribution of this sample, 
which represents an average look-back 
time of $\sim$5 Gyr, is offset from the Tully-Fisher relation (TFR)
of local low--mass spirals,
whereas the distant high--mass spirals are compatible with the local TFR.
Taking the magnitude--limited character of our sample into account, we show that
the slope of the local and the intermediate-$z$ TFR would be in compliance
if its 
scatter decreased by more than a factor of 3 between $z \approx 0.5$ and
$z \approx 0$. 
Accepting this large evolution of the TFR scatter, we hence find no strong
evidence for a mass- or luminosity-dependent 
evolution of disk galaxies. On the other hand,
we derive stellar $M/L$ ratios that indicate a luminosity-dependent
evolution in the sense that distant low--luminosity disks have much lower 
$M/L$ ratios than their local counterparts, while high--luminosity disks barely
evolved in $M/L$ over the covered redshift range. This could be the manifestation
of the ``downsizing'' effect, i.e. the succesive shift of the peak of 
star formation from
high--mass to low--mass galaxies towards lower redshifts.
This trend might be canceled out in the TF diagram due to the 
simultaneous evolution of multiple parameters.
We also estimate the ratios between stellar and total masses, finding that these
remained constant since $z=1$, as would be expected in the
context of hierarchically growing structure.

\end{abstract}


\section{Introduction}

The concept of Cold Dark Matter (CDM) and its prediction of hierarchical structure
growth has become an astrophysical paradigm in the last decade.
Observations of the Cosmic Microwave Background 
\citep[e.g.][]{Spe03} or the Large Scale Structure 
\citep[e.g.][]{Bah99} strongly support the idea that the vast majority of
matter is non--baryonic, non--luminous and interacts only gravitationally.
In this picture, gas and stars are embedded in Dark Matter halos, and
low--mass systems were the first virialized structures in the early cosmic
stages, followed by an epoch of accretion and merger events during which
larger systems were successively built up. 
Disks are destroyed during the frequent mergers of similar-mass galaxies at
$z > 1$ but can regrow via accretion events at lower redshifts 
\citep[e.g.][]{Aba03}.

An important tool for the quantitative test 
of the predictions from numerical simulations based on the hierarchical
scenario are scaling relations that link galaxy parameters.
Within the last years, several studies focused on the evolution of such
relations up to redshifts $z \approx 1$~-- corresponding to more than half the
age of the universe~-- or even beyond. Many analyses utilized 
the Tully--Fisher Relation (TFR) between the maximum rotation velocity $\vm$ and
the luminosity or stellar mass
\citep[e.g.][]{Vog01, Zie02, Mil03, Boe04, Bam06, Con05, Flo06, 
Nak06, Wei06, Chi07, Kas07}.

Studies focusing on the stellar mass TFR 
consistently have found no evidence for evolution over the past $\sim$\,7 Gyr
\citep[e.g.][]{Con05, Kas07}.
On the other hand, analyses exploiting the rest-frame 
$B$-band TFR~---
which is less sensitive to the stellar mass than to the
fraction of young stars~--- yielded quite discrepant results.
This is somewhat suprising since most recent TFR studies comprise  
$\sim$\,100\,objects or more.
Some authors derived much higher luminosities of spiral galaxies
in the past \citep[e.g.][]{Boe04, Bam06, Chi07}
while others find only a very modest evolution in luminosity 
\citep[e.g.][]{Vog01, Flo06}.
There also still is debate on whether the $B$-band TFR changes its slope 
with redshift which would imply an evolution depending on $\vm$ and hence
total galaxy mass. In \citet[][hereafter B04]{Boe04}, 
we found that low-mass spirals at $\langle z \rangle \approx 0.5$ were
brighter by up to several magnitudes than in the local universe, while
high-mass spirals barely evolved in $M_B$ at given $\vm$. To the contrary,
\citet{Wei06} observe a stronger brightening in the high-mass than in the
low-mass regime towards larger redshifts.
From the theoretical side, the situation is similarly unclear: in some
simulations, the $B$-band TFR slope remains constant with lookback time and
only the magnitude zero point changes 
\citep[e.g.][]{Ste99, Por07}, while also an evolution in slope is predicted by
other authors
\citep[e.g.][]{Boi01, Fer01}.

A growing number of observational studies based on other methods than the TFR 
have pointed towards a mass-dependent evolution of distant galaxies, either in 
terms of their colors 
\citep[e.g.][]{Kod04}, mass--to--light ratios \citep[e.g.][]{vdW05},
or average stellar ages \citep[e.g.][]{Fer04}.
These results
indicate that the global stellar populations of distant high--mass galaxies are
on average
older than those of distant low--mass galaxies, similar to what has been found
in the local universe \citep[e.g.][]{BdJ00}.
Since small galaxies are understood as
ancient building blocks within the framework of hierarchical structure growth,
and in turn should have an older stellar content than larger systems 
formed more recently, these observations are at variance with the
straightforward expectation.
They can be interpreted such that the peak of star formation 
shifts from high--mass to low--mass galaxies with growing cosmic age, a
phenomenon that was termed ``downsizing'' by \citet{Cow96}.
It is however a non-trivial question what kind of imprint ``downsizing'' has
on the TFR since the luminosity-$\vm$ diagram most likely
probes the evolution of various galaxy properties simultaneously,
including stellar $M/L$ ratio, gas mass fraction, dust content etc.

In this paper, we report on an extensive observational study of
spiral galaxy evolution over the last 7\,Gyr.
The new sample holds 124 intermediate-$z$ disk galaxies with reliable $\vm$ 
measurements, roughly doubling the size of our previous data set from B04.
The analysis given therein was limited to ground-based imaging,
while we here can rely on our high-resolution imaging performed with the
Advanced Camera for Surveys of the \emph{HST}. This allows a much more accurate
determination of disk inclination angles, scale lengths etc., and in turn
a more robust estimate of the maximum rotation velocities.
Extending the approch used in B04, we will analyse the evolution not only of the
TFR but also of the stellar mass-to-light ratios and the stellar mass fractions.
We stress that this is the first refereed publication on the total sample;
only some brief excerpts were shown in 
\citet[][]{Zie06}.

The structure of this paper is as follows. In \S2, we describe the target
selection, spectroscopic data reduction and redshift determination. \S3
briefly introduces the \emph{HST/ACS} imaging and morphological
analysis. Details on the derivation of the intrinsic maximum rotation
velocities and luminosities are given in \S4. We present and discuss our results
in \S5, followed by a short summary in \S6.
Throughout this article, we will assume a flat cosmology with
$\Omega_\lambda=0.7$, $\Omega_m=0.3$ and $H_0=70$\,km\,s$^{-1}$\,Mpc$^{-1}$.

\section{VLT spectroscopy}

We selected our targets utilizing the multi-band photometric surveys
in the FORS Deep Field 
\citep[FDF,][]{Hei03}
and William Herschel Deep Field 
\citep[WHDF,][]{Met01}. Note that the sample used in B04 comprised only FDF data.
The following criteria were applied: a) total apparent brightness $R<23^m$, 
b) spectrophotometric type ``later'' than Sa, based on a photometric redshift
catalogue in the case of the FDF 
\citep[see][]{Ben01}
or color-color diagrams for the WHDF objects 
(no photometric redshifts were available for these), adopting the 
evolutionary tracks given in \citet[][]{Met01},
c) disk inclination angle $i>40^\circ$ and misalignment angle between apparent
major axis and slit direction $\delta < 15^\circ$. The two latter constraints
were chosen to limit the geometric distortions of the observed rotation
curves. Apart from these limits, our selection was morphologically ``blind''.
No selection on emission line strength was performed.

Using the FORS 1 \& 2 instruments of the VLT in multi-object spectroscopy (MOS)
mode, we took spectra of a total of 220 late-type
galaxies between September 2000 and
October 2002. 
In the MOS masks, we also included a total of $\sim$\,40 early-type
  galaxies for a different study \citep[see][]{Zie05}. In the case that a
  target initially selected as an E/S0 galaxy turned out to have the spectrum
  of an early-type spiral (Sa/Sab), it was included in the disk galaxy sample
  a posteriori.
  Such a strategy helps to avoid selection biases against high $M/L$ ratios.

We used a fixed slit width of 1.0\,arcsec which resulted in spectral
resolutions of $R \approx 1200$ for the FDF observations, where
the grism 600R was used, and $R \approx 1000$ in the WHDF with the grism 
600RI, respectively. Each setup was exposed for a total of
2.5\,hours under seeing conditions between 0.43\,arcsec and 0.92\,arcsec FWHM,
with a median of 0.76\,arcsec. The spatial sampling in the final spectra was
0.2\,arcsec/pixel (FDF) and 0.25\,arcsec/pixel (WHDF, after the FORS\,2
CCD upgrade).

The data reduction was conducted on single extracted spectra. 
All reduction steps were performed on the
individual exposures before they were finally combined.
The typical rms of the dispersion relation fitted for wavelength calibration
was 0.04\,\AA.
For 202 of the galaxies, a redshift determination and spectral classification
was feasible. The objects range from $z=0.03$ to $z=1.49$ with a median 
$\langle z \rangle = 0.43$ (the redshift distribution of the final sample
entering the Tully--Fisher analysis is given in \S4).

\section{HST/ACS imaging}

The HST/ACS observations  
were carried out during cycles 11 and 12.
To cover the $6 \times 6$\,arcmin$^2$ sky areas of the FORS and William
Herschel Deep Fields, $2\times2$ mosaics
were taken with the Wide Field camera that has a field--of--view of
$\sim$\,$200\times200$\,arcsec$^2$ and a pixel scale of 0.05\,arcsec. 
4 visits each with a total
exposure time of 2360\,s (FDF) and 2250\,s (WHDF) through the F814W filter were 
used per quadrant. Each visit was split into two exposures.
We kept the resulting frames from the ACS pipeline reduction 
(including bias subtraction, flatfielding and distortion correction) and
used a filtering algorithm to combine the two exposures of each visit
and remove the cosmics. 

The structural parameters of the galaxies were derived with the GALFIT package 
\citep{Pen02}. 
For convolution of the model profiles, a mean Point Spread Function was constructed using
$\sim$\,20 unsaturated stars with $I_{814}<23$ which
were normalized to the same central flux and median--averaged.
An exponential profile was used to fit the galaxies' disk components, while a
S\'ersic profile was taken to model an additional bulge, where detectable.
Note that for the analysis presented here, the most important parameters are
the inclination, position angle and scale length of the disk. The best-fitting 
bulge parameters were only used for a raw morphological classification.

We found significant bulge components in only 46 out of 124 galaxies  
that were reliable for $\vm$ determination (see next section), 
spanning the range $0.01 \le B/T \le 0.53$ (median $\langle B/T \rangle =  0.15$).
In terms of a visual classification within the Hubble scheme, our sample comprises
all types from Sa spirals to irregulars.

\section{\label{vmax}Tully--Fisher Parameters}

To extract rotation curves (RCs) from the 2D spectra, Gaussian fits were applied to
the emission lines stepwise perpendicular to the direction of
dispersion. An averaging boxcar of 3 pixels widths, corresponding to
0.6\,arcsec, was used to increase the
$S/N$. For very weak lines, this boxcar size was increased to 5 pixels.
In cases of multiple usable emission lines for a given object,
the RC with the largest spatial extent and highest degree of symmetry was used
as reference. RCs from different emission
lines agreed within the errors (at least in terms of the derived maximum rotation
velocities)
for the majority of the objects. Most of the ``reference'' RCs were extracted from 
the [O {\small II}] or [O {\small III}] lines, a smaller fraction from
H$_\alpha$ or H$_\beta$ (for a vast number of example rotation curves, please
see B04).

Due to the small apparent sizes of the spirals in our sample, the observed
rotation curves were heavilly blurred.  
At redshifts of $z \approx 0.5$, the apparent scale lengths are of the
same order as the slit width (one arcsec for the entire data set) 
and the seeing disk. 
The maximum rotation velocity hence cannot be
derived ``straightforward'' from the outer regions of the observed
rotation curves.

To account for the blurring effects, 
we generated synthetic rotation velocity
fields according to the measured
disk inclination, position angle and scale length
of a given object. We assumed an intrinsic linear rise of the rotation
velocity at small radii
which turns over into a regime of constant rotation velocity
at large radii, with the turnover radius computed from the
scale length of the emitting gas 
(for details on this prescription as well as 
tests of various rotation curve shapes, 
we again refer to B04).
Using the position angle and disk inclination derived with GALFIT,
a 2D rotation velocity field then was generated.
After weighting with the luminosity profile and
convolution with the Point Spread Function,
a ``stripe'' was extracted from the velocity field that corresponded to
the orientation and width of the given slitet used for spectroscopy.
This simulation
yielded a \emph{synthetic} rotation curve which was fitted to 
the corresponding observed curve to derive the intrinsic value of $\vm$.

Following this approach, $\vm$ values could be determined for 124 spirals
in our sample (hereafter, we will refer to this kinematic data set as the 
TF sample). 
78 galaxies were rejected, partly due to disturbed kinematics, which
could introduce systematic errors in the TF analysis.
Other objects were not suitable due to low $S/N$,
solid-body rotation or the lack of significant rotation
within the measurement errors.
The TF sample galaxies span the redshift range $0.05<z<0.97$ (median
0.45, corresponding to a lookback time of 4.7\,Gyr) 
and comprise 19 objects with spectrophotometric type Sab,
65 Sc spirals and 40 Sdm galaxies. $\vm$ values fall in the range
22\,km/s $ < \vm<$ 450\,km/s (median $\langle \vm \rangle$ = 135\,km/s).

The computation of the luminosities benefitted strongly from the
multi--band imaging of our targets. 
Depending on the redshift of a given object,
the filter which best matched the rest--frame $B$-band was used to
compute the absolute magnitude $M_B$. 
This way, systematic $k$-correction errors
due to dependence on the SED are very small ($<$0.1\,mag for all types
and redshifts in our data set).
Intrinsic absorption was corrected following 
\citet{TF85}. 
In this approach, the amount of absorption only depends on the
disk inclination. For testing purposes, we alternatively applied the
inclination-- \emph{and} mass--dependent correction factors given by 
\citet{Tu98}.
However, this did not change the results presented in the following.
It only is crucial that any distant sample is corrected for internal
absorption \emph{in the same way} as the local sample it is compared to.
Since we use the  data from 
\citet{PT92} as local reference here, who adopted the
Tully \& Fouqu\'e approach, we will keep this convention in the following.
The TF galaxies have absorption-corrected
absolute magnitudes between $M_B=-16.8$ and $M_B=-22.7$
(median $\langle M_B \rangle$ = $-$20.3).
The 78 galaxies with RCs that were not reliable for a $\vm$ determination have
slightly lower luminosities, with a range $-22.4 < M_B < -14.8$ and a median
value of  $\langle M_B \rangle$ = $-$19.8.

\begin{figure}
\includegraphics[bb=55 37 560 611,angle=270,scale=0.8]{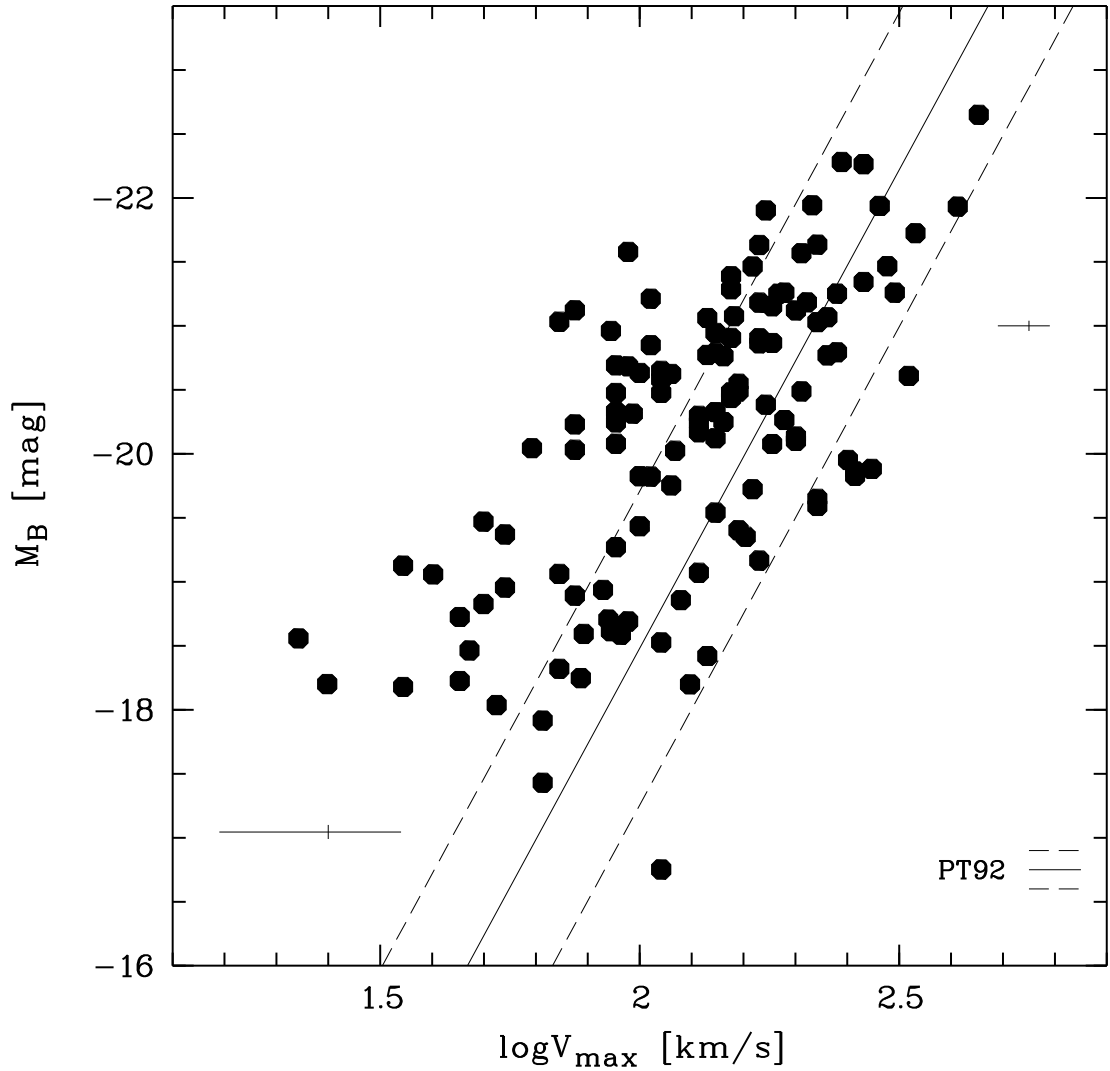}
\caption
{The distant FDF \& WHDF galaxies at 
$\langle z \rangle = 0.45$ compared to the local Tully--Fisher relation
as given by Pierce \& Tully (1992, solid line; 
the dashed lines denote the 3\,$\sigma$ limits). 
The distant spirals are
systematically overluminous for their values of $\vm$. Typical error bars
for the high--mass and low--mass regime are indicated in the upper right and
lower left corner, respectively.
\label{tfplot}}
\end{figure}

In Fig.~\ref{tfplot}, we compare our sample to the local Tully--Fisher relation from 
\citet[][PT92 hereafter]{PT92}.
On average, the FDF \& WHDF spirals are more luminous than
their local counterparts by 
$\langle \dmb \rangle = -0.84^m$,  which could be due to younger stellar populations,
i.e.~lower $M/L$ ratios of the distant galaxies. 
This interpretation is supported by the fact that the amount of the distant
galaxies' brightening $\dmb$ increases with redshift. A linear
$\chi^2$-fit yields 
$\dmb=(-1.22\pm0.40)\,z-(0.02\pm0.20)$[mag]. 
Here, we took
the error propagation of $\vm$ errors as well as 
the typical errors for the local spirals into account.
This luminosity evolution is in agreement with the findings by, 
e.g., \citet{Mil03}, 
but exceeds the very modest TF offsets derived by \citet{Vog01}.

\section{Discussion}

The luminosity--rotation velocity distribution of our sample  indicates that
the offsets
from the local Tully--Fisher Relation change not only with redshift (see
above)
but also with mass:
while the high--mass spirals are in relatively good agreement with the local
TFR, the low--mass spirals are overluminous by up to several magnitudes at
given $\vm$ (cf.~Fig.~\ref{tfplot}). Using a 100 iteration bootstrap bisector fit, 
we find an intermediate-redshift TFR of
$M_B=-(4.27\pm0.30)\,\log\vm-(11.18\pm0.65)$, 
which is a significantly shallower slope 
($a=-4.27$) than locally, where $a=-7.48$ is observed \citep[][]{PT92}; 
we emphasize that this published value is in good
agreement with a bisector fit to the local sample which yields $a=-7.57 \pm 0.38$. 
Note that the bisector fitting method~-- which is a combination of a ``forward'' 
and an ``inverse'' TF fit~--
can be only weakly affected by potentially correlated errors
in $\dmb$ and $\vm$ suspected by 
\citet{Bam06}.

We reported earlier on a potential mass--dependent
luminosity evolution with an analysis that was limited to 
the FDF sub--sample and ground--based structural parameters 
\citep[][B04]{Zie02}.
Therein, we ruled out a variety of systematic errors that potentially could
bias the observed distant TFR slope.
Firstly, we tested whether 
tidally induced star formation in close galaxy pairs affects our sample,
but found that the TF distributions of pair candidates and isolated galaxies
are consistent.
Secondly, our results were robust against the use of different prescriptions 
for the intrinsic absorption correction.
Thirdly, we sub-divided the FDF sample according to the rotation curve quality
in terms of radial extent and symmetry. A re-analysis using only 
high-quality rotation curves 
confirmed the shallower distant TF slope, showing that our
findings are not induced by perturbated or truncated rotation curves.

We also tested whether a slope change could be mimicked by
an incompleteness effect arising from the apparent magnitude limit in 
our target selection. Towards higher redshifts, such a limit corresponds to 
higher luminosities and, in turn, higher masses. A fraction of the
low--luminosity, low--mass (slowly rotating) spiral population is therefore
missed in the selection process, while the  low--mass galaxies that \emph{are}
selected might preferentially be located at
the high--luminosity side of the relation. 

\begin{figure}
\includegraphics[bb=55 37 560 611,angle=270,scale=0.8]{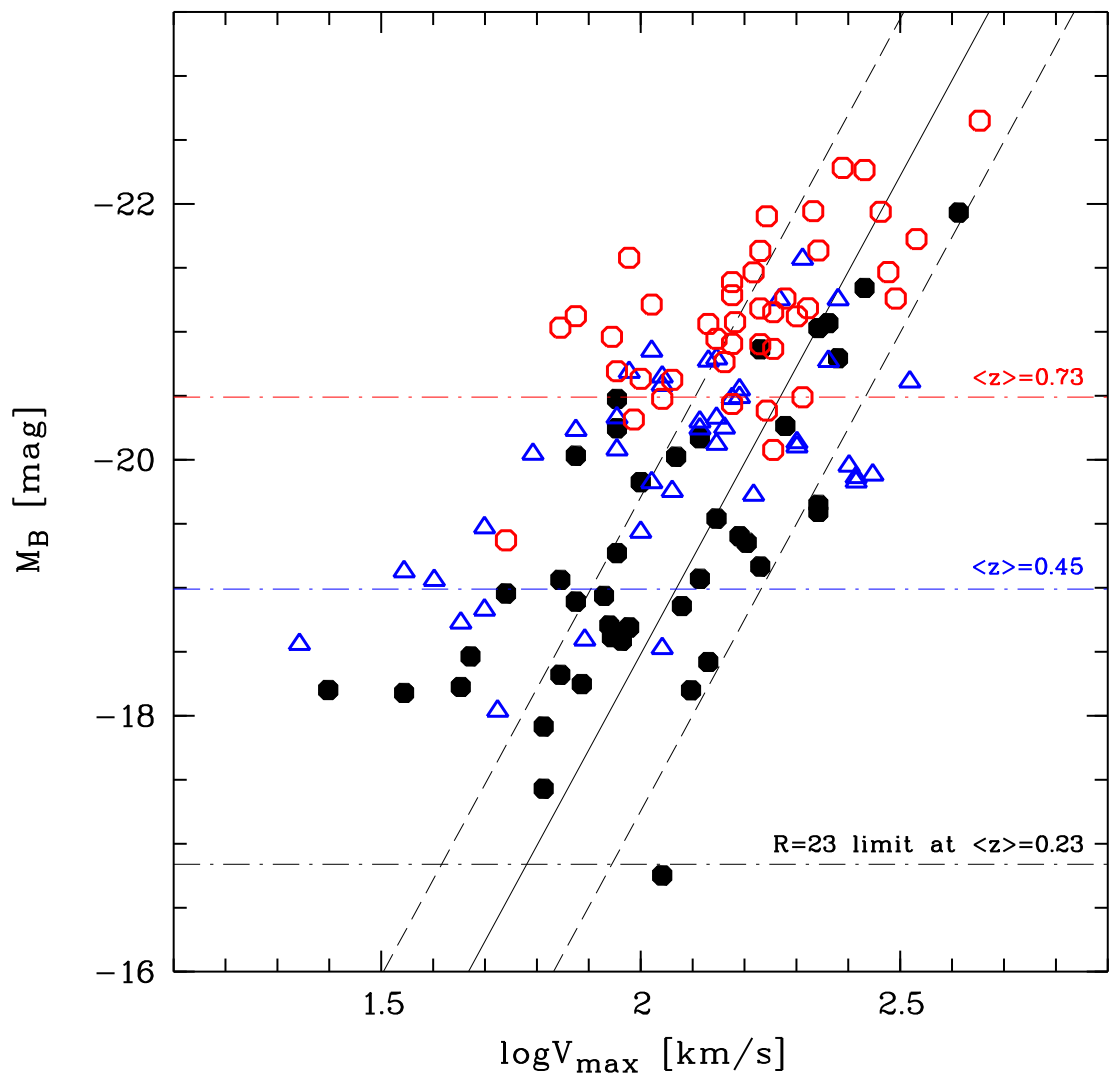}
\caption{
The FDF \& WHDF galaxies compared to the local TFR as given by Pierce \& Tully 
(1992, solid line; the dashed lines denote the 3\,$\sigma$ limits). 
The distant sample is divided into three equally-sized sets with median
redshifts $\langle z \rangle = 0.23$ (solid circles),  
$\langle z \rangle = 0.45$ (open triangles) and $\langle z \rangle = 0.73$
(open circles), respectively.
The horizontal dot-dashed lines depict the limit in $B$-band absolute
magnitude corresponding to our selection on apparent brightness $R < 23^m$.
An SED corresponding to Hubble type Sc has been assumed here.
\label{zbins}
}
\end{figure}

To test this effect on the new, full sample, we split it into
three equally-sized redshift bins corresponding to median redshifts
$\langle z \rangle = 0.23$, $\langle z \rangle = 0.45$ and
$\langle z \rangle = 0.73$, respectively (see Fig.~\ref{zbins}).
In all redshifts bins, at given $\vm$ the distant galaxy sample shows an 
overluminosity of the low-mass galaxies compared to the local TFR, while
the distributions are similar at the high-mass end. 
Using bootstrap bisector fits, we find significant deviations from the local
TFR slope in all redshift bins: 
$a=-4.18\pm0.35$ at $\langle z \rangle = 0.23$,
$a=-2.77\pm0.56$ at $\langle z \rangle = 0.45$ , and 
$a=-3.00\pm0.50$ at $\langle z \rangle = 0.73$, respectively.
Also shown in Fig.~\ref{zbins} 
are the absolute magnitudes to which our selection criterion $R<23^m$
corresponds at the median redshift of each sub-sample (dot-dashed lines).
To derive
these limits, we computed the $k$-correction using a synthetic spectrum of type
Sc by \citet[][]{Moe01}. Note that the
$k$-korrection for transformation from $R_{\rm obs}$ to $B_{\rm rest}$
is only weakly depending on SED type at redshifts $0.3 \lesssim z \lesssim 0.8$.
To achieve consistency, we corrected for intrinsic absorption at a disk inclination
angle of $i=60^\circ$, which is the average of our data set.
It is evident from Fig.~\ref{zbins} that the magnitude limit of our survey
affects the 
covered luminosity ranges in the intermediate- and high-redshift bin. In the 
low-redshift bin ($\langle z \rangle = 0.23$), the lack of galaxies with
$M_B \gtrsim -18$ seems less likely to be induced by the $R<23^m$ selection.
In the following, we aim at a quantitative estimate of the impact of sample
incompleteness on the TF analysis.

The key factor for the strength of this selection effect is the scatter of the TFR, 
which has a value of $\sigma_B=0.41^m$ in the $B$-band of the local PT92 sample. 
In a previous analysis \citep[][]{Zie02},
we assumed that this scatter increases by a factor of 1.5 between $z \approx 0$ and
$z=0.5$ due to, e.g., a broader distribution in star formation rates at
earlier comic stages. Here, we will use a different approach by testing
\emph{how strongly the TF scatter would have to evolve over the past
$\sim$5\,Gyr for the $\vm$--dependent TF offsets to be attributed to
an incompleteness effect}. 

We hence assumed that the slope of the TFR
remained constant over the redshift range under scrutiny here, and that its
scatter has been larger at earlier cosmic times. Similar to the technique
described by \citet{Gio97}, 
a Schechter luminosity function (LF) form was fitted to the observed luminosity 
distribution, with
the characteristic luminosity $M^\ast$ and the space density $\phi ^ \ast$ as
free parameters. At the faint end of the luminosity function, where our sample
is incomplete, a slope of $\alpha = -1.2$ was adopted, which is a
typical value found in studies of the $B$-band LF at intermediate redshifts
\citep[e.g.][]{Gab04,Gia05}.
We could not determine the faint end slope directly from the data since
the luminosity distribution of our sample peaks at $M_B \approx -20$ and falls
off towards fainter magnitudes due to incompleteness.
%Within magnitude bins of 0.75\,mag width, 
The ratio between the observed LF
and the best-fit Schechter LF was computed and expressed as an incompleteness
function $y=f(M_B)$ that ranged from $y=1$ in the case of 100\% completeness to $y=0$
in the case of 0\% completeness. On the basis of the local TFR, $y=f(M_B)$ 
was then converted to $y=f(\vm)$. This was done because the distribution in
$\vm$ is much less affected by the apparent magnitude limit than the
distribution in $M_B$, hence the incompleteness bias was computed as a function
of $\vm$, not as a function of $M_B$.

To derive the impact of the sample incompleteness, an \emph{unbiased} TFR 
of the form $M_B=a \log \vm + b$ was
assumed with a fixed slope $a=-7.48$.
The TFR intercept $b$ was determined
implicitely from the best-fit value of the charateristic luminosity $M^\ast$.
We stress again that the purpose of this approach was only to test whether
a time-independent TF slope could be consistent with our data set.
For each object in the TF sample with an \emph{observed} maximum
rotation velocity $\vm$, the incompleteness function $y=f(\vm)$ was taken as the
probability that the given galaxy would enter the observed TF sample with an
observed, biased absolute magnitude $M_B^b=a \log \vm + b \pm \sigma(\vm)$.
As observed in the local universe \citep[e.g.][]{Gio97}, the scatter $\sigma(\vm)$
was assumed to be larger for lower mass galaxies 
(at, e.g., $\vm = 100$\,km/s, the scatter was taken to be a factor
of 1.6 larger than at $\vm = 300$\,km/s)
with an average
value satisfying $\langle \sigma(\vm) \rangle = \sigma_B$.
This computation of $M_B$ including a TF scatter 
was iterated 800 times for every object, yielding a simulated,
\emph{biased} TF relation.
The difference between the absolute magnitude $M_B=f(\vm)$ of the un-biased
TFR and the simulated absolute magnitude $M_B^{\rm b}=f(\vm)$ of the biased
TFR was taken as the \emph{correction factor} to de-bias our TF sample.
Note that the de-biased magnitudes $M_B^{\rm db}$ are fainter
than the observed values 
and that the TFR becomes steeper after de-biasing.

These computations were performed in three variants using 
i) the local scatter $\sigma_B$, ii) a 2\,$\times$ local scatter
and iii) a 3\,$\times$ local scatter for the intermediate-$z$ TFR. The
corresponding de-biased TF samples were fitted with bootstrap bisector fits.
To account for the fact that the distant galaxies cover a range in
redshift, corresponding to various lookback times, the multiplicating factor
for introducing the scatter in the simulated distant TFR was not
kept constant for the whole sample but computed individually for each object
in such a way that the \emph{average} scatter was either equal to $\sigma_B$,
$2\,\sigma_B$, or $3\,\sigma_B$.

Using this approach, we found that the un-biased distant TFR slope would be
$a=-4.88 \pm 0.29$ if the distant 
TF scatter was the same as locally. If the scatter doubles between
$z=0$ and $z \sim 0.5$,
the incompleteness effect becomes stronger and the de-biased distant TFR slope
would be 
$a=-6.09 \pm 0.28$. Assuming a three times larger scatter than locally we
found a de-biased distant slope of $a=-7.64 \pm 0.27$. 
However, the local TFR is \emph{also} affected by a magnitude limit. 
The de-biased local TF bisector fit slope we find is 
$a=-8.02\pm 0.41$ (of course, 
the unchanged local TF scatter of $\sigma_B=0.41^m$ was used here). 
We hence conclude that the distant and local TFR would be consistent in terms of an
incompleteness effect only
\emph{if the TF scatter evolved by more than a factor of 3 over the past 
$\sim$5\,Gyr}. This would be in agreement with the observed distant scatter which is
$\sigma_B \approx 1.2$\,mag.
On the other hand, this might be an overestimate since the free-fit scatter (i.e.,
with a slope $a=-4.27$, see beginning of this section) is only $\sim 0.9$\,mag.

\begin{figure}
\includegraphics[bb=66 41 553 622,angle=270,scale=0.8]{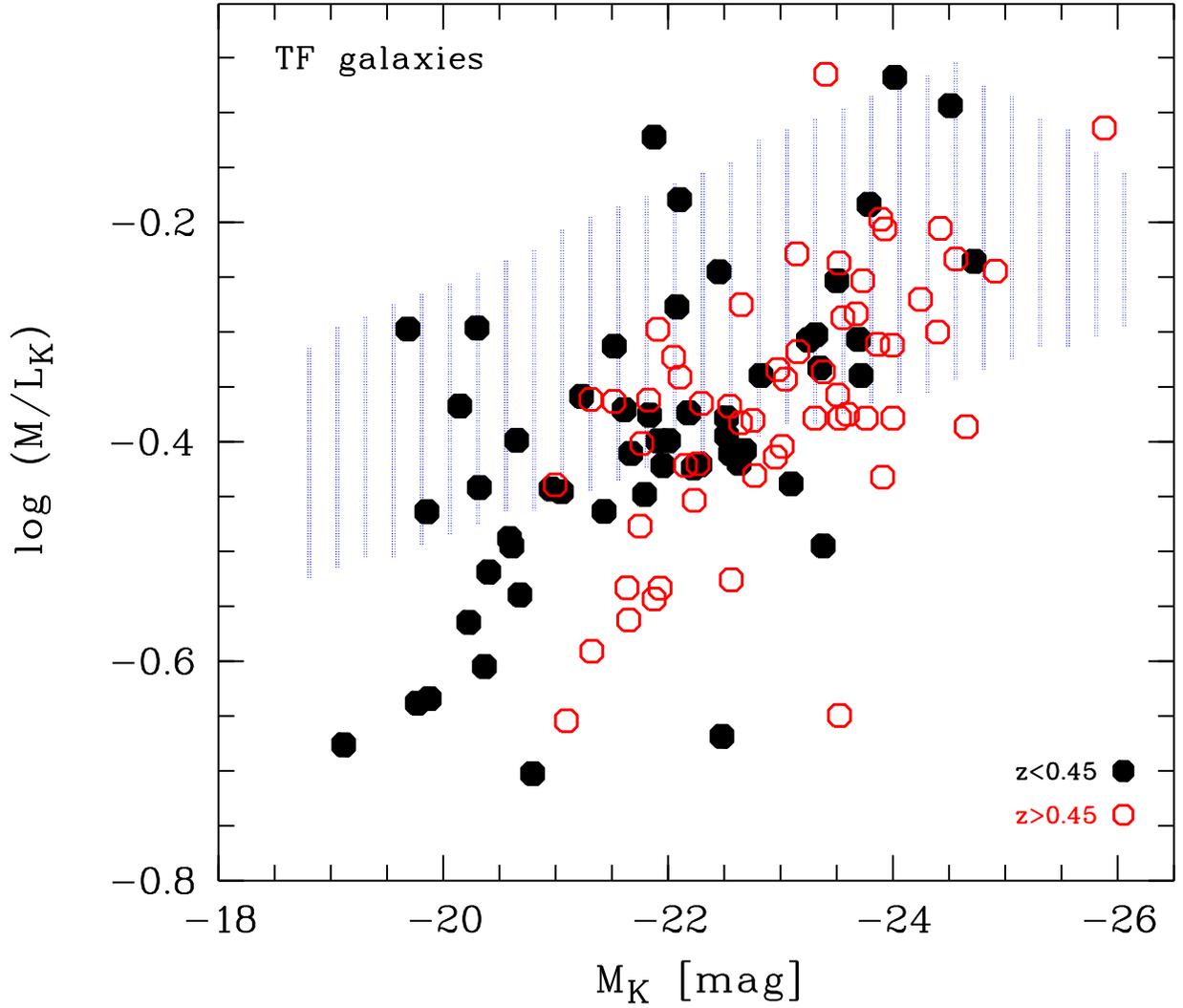}
\caption{
Stellar mass-to-light ratios of our distant TF sample galaxies at
$0.1< z<0.45$ (filled circles) and $0.45<z<1.0$ (open circles), 
compared to the parameter range
covered by present-day spirals (shaded area) from \citet{BdJ01}. The data
indicate a stronger evolution in $M/L$ for low-luminosity galaxies.
\label{mlk1}}
\end{figure}

These results imply that the differences between the distant and local
TF distributions can at least in part be attributed to an incompleteness effect.
However, the TFR directly traces only the overall luminosities of a
galaxy sample,  not the properties of the stellar populations.
To perform a more detailed analysis, we converted 
the rest--frame, absorption-corrected $B-R$ colors of the galaxies in our
sample into $K$-band $M/L$ ratios following \citet{BdJ01}. 
In Fig.~\ref{mlk1}, we show
the $M/L_{\rm K}$ ratios versus measured
$K$-band absolute magnitudes of the distant TF
spirals in comparison to local disk galaxies.
We have split the distant galaxies into two redshift bins covering 
$z<0.45$ (filled symbols) and $z>0.45$ (open symbols), respectively.
It is evident that the distant TF galaxies on the average have
lower mass-to-light ratios at a given luminosity than their present-day
counterparts (shaded area; note that several of the distant galaxies are at 
the \emph{upper} limit of the local distribution, 
indicating that our selection did not miss disks with high $M/L$ ratios). 

\begin{figure}
\includegraphics[bb=66 41 553 622,angle=270,scale=0.8]{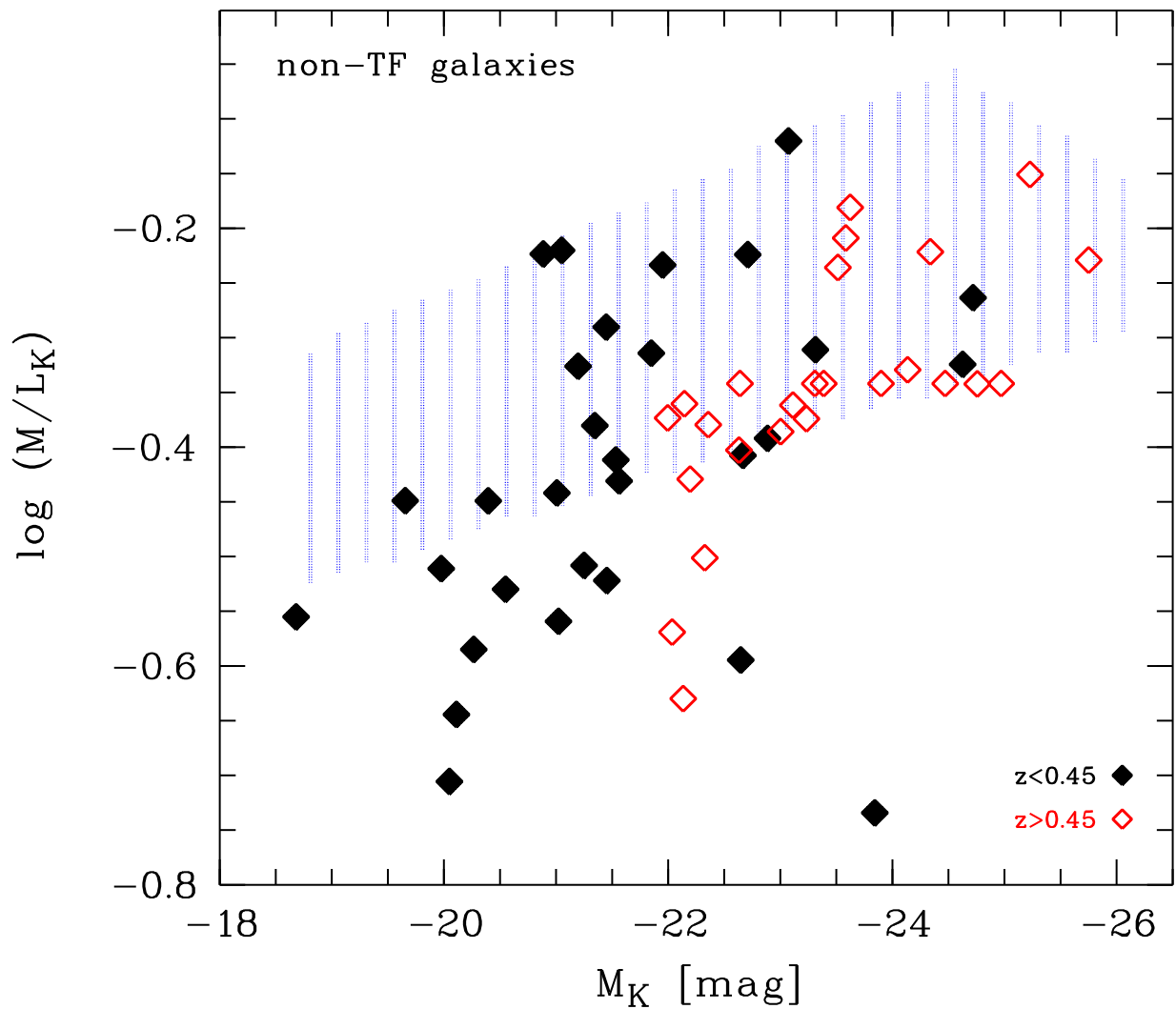}
\caption{
Stellar mass-to-light ratios of the distant galaxies which were not usable for
a determination of $\vm$, splitted into two redshift bins for
$0.1< z<0.45$ (filled symbols) and $0.45<z<1.0$ (open symbols), respectively.
The parameter range covered by present-day spirals from \citet{BdJ01} is
indicated by the shaded area. Similar as for the TF objects (see
Fig.~\ref{mlk1}), the data indicate a stronger evolution in $M/L$ for low-luminosity galaxies. 
\label{mlk2}}
\end{figure}

Moreover, there is a dependency on luminosity: low-luminosity
galaxies seem to evolve stronger in $M/L_{\rm K}$ than high-luminosity
galaxies. 
One possible explanation for this could be a larger fraction of young stars in
the distant low-luminosity galaxies than high-luminosity galaxies.
Since the $K$-band luminosity is a good tracer of stellar mass,
this would imply higher luminosity-weighted stellar ages towards
higher stellar masses.

For comparison, 
we also computed the luminosities and mass--to--light ratios of the disk
galaxies which were not included in the TF sample due to kinematic
disturbancies, low S/N ratios of their emission lines or lack of significant
rotation. These galaxies are shown in Fig.~\ref{mlk2}, divided into two
$z$-bins as in Fig.~\ref{mlk1}.
The non-TF galaxies show a similar trend as the distant TF spirals, i.e.~a stronger
evolution of the $M/L$ ratios 
in the low--luminosity regime with respect to the high--luminosity regime.
Note that there is no evidence for very blue, luminous disk galaxies that 
have been excluded from the TF analysis (in principle, such a population of luminous disks with
blue colors and irregular kinematics could be expected within the framework of 
the hierarchical scenario, e.g.~due to high--mass spirals which have recently
undergone merger events).

Could the observed evolution of the $M/L$ ratios 
indicate a difference in stellar ages between high-- and
low--luminosity disks in the distant universe?
Indeed, this interpretation gains support from an analysis  
of the broad-band colors of galaxies at $z>0.5$ from our data set 
with single-zone models of chemical enrichment, which
yielded evidence for 
a dependence of the mean stellar ages on $\vm$ (and hence total mass): the
high-mass galaxies have older stellar populations than the low-mass ones 
\citep[see][]{Fer04}. 
This indication for an anti-hierarchical evolution of the baryonic component
of galaxies 
(``downsizing'') has also been found in other studies of distant galaxies 
\citep[e.g.][]{Kod04, vdW05}.
It is possible that the various evolutionary effects coming into play between
the earlier and the local universe~--
the evolution of the stellar populations, gas mass fractions, dust content etc.~--
balance each other in such a way that the downsizing phenomenon is \emph{not} 
reflected in a significant differential evolution of the Tully--Fisher Relation.

The lack of hierarchical evolution in terms of the stellar populations raises
the question whether a hierachical evolution of the Dark Matter Halos~---
the most fundamental prediction of ($\Lambda$)CDM cosmology~--- can be
established with the data.
To test this, we will use a similar approach as 
\citet{Con05} by focussing on the evolution of the ratio between stellar and
total mass since redshift $z=1$. 
Based on the $K$-band mass-to-light ratios,
we transformed the absolute $K$
magnitudes into stellar masses $M_\ast$. Total masses were estimated from
the disk scale lengths and maximum rotation velocities adopting the results of
\citet{vdBo02}.
Our sample covers the ranges 
$2.0 \cdot 10^8\,M_\odot < M_{\ast} < 3.7 \cdot 10^{11}\,M_\odot$ 
(median $\langle M_{\ast} \rangle = 9.0 \cdot 10^{9}\,M_\odot$) 
in stellar mass and
$2.5 \cdot 10^9\,M_\odot < M_{\rm vir} < 5.2 \cdot 10^{12}\,M_\odot$ 
(median $\langle M_{\rm vir} \rangle$~=~1.1 $\cdot 10^{11}\,M_\odot$) in total mass.

\begin{figure}
\includegraphics[bb=62 39 560 602,angle=270,scale=0.8]{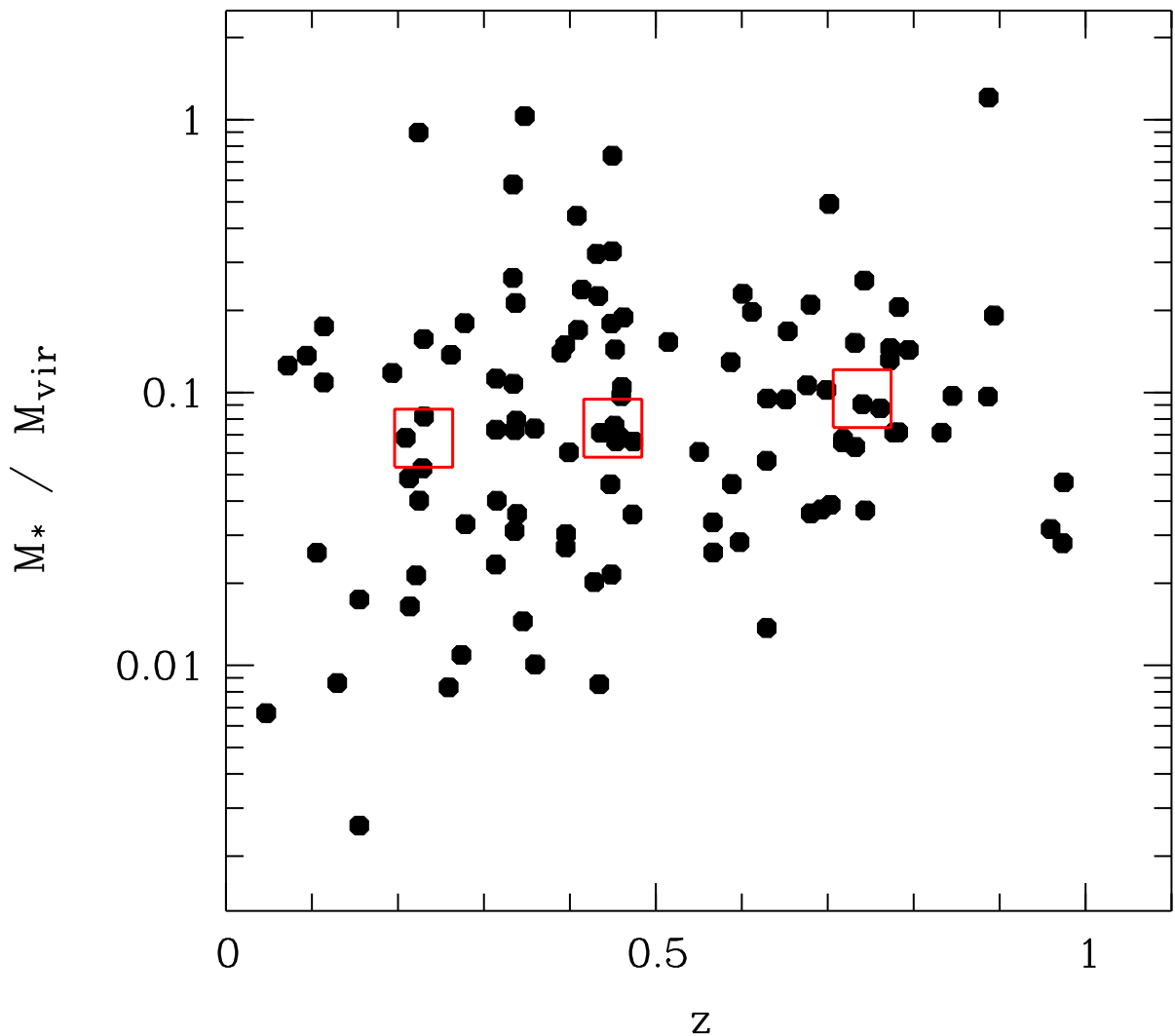}
\caption{
The stellar--to--total mass fractions of the 110 galaxies from our sample with determined
$\vm$ and available $K$-band photometry. The three large open squares show the
median mass fraction within three redsift bins containing 36-37 galaxies each.
The lack of a stellar mass fraction increase towards lower redshift favours
a hierarchical buildup of the galaxies (see text for details).
\label{frac}}
\end{figure}

In Fig.~\ref{frac}, we show the ratio between stellar and total mass as a function of
redshift. If we sub--divide our sample into three redshift bins with
36 to 37 galaxies in each bin, we find median stellar mass fractions of
0.068, 0.074 and 0.095 
at redshifts $z<0.35$, $0.35<z<0.6$ and $z>0.6$, respectively.
We hence observe a very slight decrease of the stellar mass fraction of
spirals between $z=1$ and the local universe.
If late-type galaxies contained all their gas by redshift unity and
only converted gas into stars since then, an increase of the stellar mass
fraction would be the result.
Instead of such a ``monolithic'' scenario, which is reliable to
describe the evolution of massive ellipticals, the data indicate that
spiral galaxies have accreted baryonic (and most probably also dark)
matter in the regime $0<z<1$, in agreement with the observational findings of
\citet{Con05} 
and with the expectation for a hierachical structure growth 
\citep[e.g.][]{Bau05}.

\section{Conclusions}

We have used VLT/FORS spectroscopy and HST/ACS imaging to construct a sample of 
220 field spiral galaxies up to redshift $z=1$.
Spatially resolved rotation curves were extracted and fitted with synthetic
velocity fields that take into account all geometric and observational effects,
like blurring due to the slit width and seeing influence. 
Using these fits, the maximum rotation velocity $\vm$ could be determined for
124 galaxies 
with an average look-back time of $\sim$5 Gyr.
We find that this sample is offset from the local Tully-Fisher Relation.
The distant low--mass galaxies are more luminous at given $\vm$ than their
local counterparts,
whereas the distant high--mass spirals are compatible with  the local TFR.
Taking the magnitude limit of our sample into account, we show that
the slope of local and distant relation would be in compliance if the 
TFR scatter decreased by more than a factor of 3 between $z \approx 0.5$ and
$z \approx 0$.
On the other hand,
the $M/L$ ratios indicate a luminosity-dependent
evolution in the sense that distant low--luminosity disks have much lower 
$M/L$ ratios than their local counterparts, while high--luminosity disks barely
evolved in $M/L$ over the covered redshift range. This could be interpreted as an
indication of the ``downsizing'' effect, 
i.e. the succesive shift of star formation from
high--mass to low--mass galaxies towards lower redshifts.
In terms of the Dark Matter Halos, we find evidence for a hierarchical evolution,
since the fraction between stellar and total mass remained roughly
constant since $z=1$.

\acknowledgments

We thank ESO for the efficient support during the spectroscopic observations 
and the FDF team for the contributions to the FDF sample analysis.
We also thank the anonymous referee for the comments and suggestions which
helped to significantly improve the paper.
We are furthermore grateful to J.~Fliri and A~.Riffeser (both USM M\"unchen) 
for the cosmic ray removal on the ACS images of the FDF and 
J.~Heidt (LSW Heidelberg) for providing WHDF pre-images.
This work was funded by the Volkswagen Foundation (I/76\,520) and
the ``Deutsches Zentrum f\"ur Luft- und Raumfahrt'' (50\,OR\,0301).


%% copy editing.  Individual instruments can be provided in parentheses,
%% after the keyword, but they will not be verified.

%Facilities: \facility{Nickel}, \facility{HST(STIS)}, \facility{CXO(ASIS)}.

\end{document}